\documentclass[preprint,superscriptaddress,nofootinbib]{revtex4-1}
\usepackage{graphicx}

\usepackage{amsmath,amssymb}

\usepackage{color}

\begin{document}
\title{Primordial black holes as a probe of strongly first-order electroweak phase transition}
\author{Katsuya~Hashino}
\affiliation{Center for High Energy Physics, Peking University, Beijing 100871, China}
\author{Shinya~Kanemura}
\affiliation{Department of Physics, Osaka University, Toyonaka, Osaka 560-0043, Japan}
\author{Tomo~Takahashi}
\affiliation{Department of Physics, Saga University, Saga 840-8502, Japan}

\begin{abstract}

Primordial black holes can be produced by density fluctuations generated from delayed vacuum decays of first-order phase transition.
The primordial black holes generated at the electroweak phase transition have masses of about $10^{-5}$ solar mass. 
Such primordial black holes in the mass range can be tested by current and future microlensing observations, such as Subaru HSC, OGLE, PRIME and Roman telescope.
Therefore, we may be able to explore new physics models with strongly first-order electroweak phase transition via primordial black holes.
We examine this possibility by using models with first-order electroweak phase transition in the standard model effective field theory with dimension 6 and 8 operators. 
We find that depending on parameters of the phase transition a sufficient number of primordial black holes can be produced to be observed by above mentioned experiments.
Our results would suggest that primordial black holes can be used as a new probe of models with strongly first-order electroweak phase transition, which has complementarity with measurements of the triple Higgs boson coupling at future collider experiments and observations of gravitational waves at future space-based interferometers. 

\end{abstract}

\preprint{OU-HET-1123}

\maketitle



\section{ Introduction }
Although the Higgs boson was discovered at the Large Hadron Collider (LHC) in 2012~\cite{Aad:2012tfa}, the Higgs sector remains unknown.
In particular, the shape of Higgs potential has not been known and a mechanism of electroweak phase transition (EWPT) is still a mystery.
The EWPT in the standard model (SM) is crossover~\cite{Dine:1992vs}.
 On the other hand, the model with Higgs sector extension can realize strongly first-order EWPT which is required by electroweak baryogenesis~\cite{Kuzmin:1985mm}, one of the promising scenarios for explaining the baryon asymmetry of the Universe (BAU).
Thus, it is crucial to reveal the nature of EWPT by experiments. 
If the EWPT is strongly first-order phase transition, 
its effect appears as a large deviation in the triple Higgs boson coupling ($hhh$ coupling) from the SM prediction value~\cite{Grojean:2004xa, Kanemura:2004ch} and gravitational waves (GWs) in the frequency range of $10^{-3}$--$10^{-1}$ Hz~\cite{Grojean:2006bp}.
Therefore, it has been discussed that measurements of the $hhh$ coupling at future collider experiments and observations of such GWs can be important probes of strongly first-order EWPT. 
The model with strongly first-order EWPT can be tested by correlation between collider and GW observations, like in Refs.~\cite{Kakizaki:2015wua, Hashino:2016rvx, Kobakhidze:2016mch, Huang:2016cjm, Hashino:2016xoj, Artymowski:2016tme, Beniwal:2017eik, Huang:2017rzf, Hashino:2018zsi, Chala:2018ari, Huang:2018aja, Bruggisser:2018mrt, Alves:2018oct, Hashino:2018wee, Ahriche:2018rao,Chala:2018opy,Alves:2018jsw,Chen:2019ebq,Alves:2019igs}.

In the early Universe, it is known that primordial black holes (PBHs) can be produced via large density contrast from various mechanisms.
One of such mechanisms is strongly first-order phase transition in the early Universe~\cite{Kodama:1982sf, Hawking:1982ga}.
Recently, there are several works discussing new mechanisms of PBH production from first-order phase transition in the early universe~\cite{Baker:2021nyl, Kawana:2021tde, Liu:2021svg, Jung:2021mku, Baker:2021sno}.
In these works, the authors are mainly interested in explaining the possibility that PBHs play a role of dark matter.

In this paper, we investigate the possibility that PBHs can be used as a probe of models with first-order EWPT. 
We employ the production mechanism proposed in Ref.~\cite{Liu:2021svg}, where the large density contrast arising from probabilistic nature of vacuum decay can produce PBHs with the mass corresponding to the era of phase transition.
For PBHs produced at the EWPT, their masses are about $10^{-5}$ solar mass. 
Such PBHs in this mass range can be tested by current and future microlensing observations, such as Subaru Hyper Suprime-Cam (HSC)~\cite{HSC}, Optical Gravitational Lensing Experiment (OGLE)~\cite{OGLE}, PRime-focus Infrared Microlensing Experiment (PRIME)~\cite{PRIME} and Nancy Grace Roman space (Roman) telescope~\cite{Roman}, from which we may be able to explore new physics models with strongly first-order EWPT via PBHs.


We first examine phase transition parameters such as released latent heat and duration of phase transition which can produce PBHs by exceeding the critical density contrast by the strongly first-order EWPT.
We then evaluate the abundance of PBHs and study whether this can be observed by current and future microlensing experiments. 
We second consider a concrete model of the Higgs sector with dimension 6 and 8 operators which can realize these phase transition parameters.
We clarify the testability of the model by using PBHs, in addition to GWs at future observations such as Laser Interferometer Space Antenna (LISA)~\cite{Klein:2015hvg}, DECi-hertz Interferometer Gravitational Wave Observatory (DECIGO)~\cite{Yagi:2011wg} and the $hhh$ coupling to be measured at future collider experiments such as High-Luminosity LHC (HL-LHC)~\cite{Cepeda:2019klc}, the International Linear Collider (ILC)~\cite{Asner:2013psa, Moortgat-Picka:2015yla, Fujii:2015jha} and so on~\cite{Asakawa:2010xj}.
Therefore, we argue that the model with first-order EWPT can be comprehensively tested.

 \section{PBH and GW productions from first-order phase transition}

We here briefly describe how to evaluate PBH abundance from the first-order phase transition according to the method in Ref.~\cite{Liu:2021svg}. 
The decay rate of the false vacuum with linear approximation\footnote{
This approximation may break down for some parameter space, and then 
the decay rate may be suitably assumed as 
$\Gamma\propto$ Exp[$-\beta_B(t^2-t_m^2)$]~\cite{Ellis:2018mja}. Such a case may require a more careful 
analysis for the phase transition and PBH production. This point will be discussed in future work~\cite{future}. 
}
is given by 
\begin{equation}
\label{eq:decay}
\Gamma(t)=\Gamma_0 e^{\beta t},
\end{equation}
 where $\Gamma_0$ is the value at the initial time $t=0$, and $\beta$ is the inverse of duration time of the phase transition which is defined as 
\begin{equation}
\label{eq:beta}
\beta = \left.\frac{1}{\Gamma}\frac{d\Gamma}{dt} \right|_{ t=t_n} .
\end{equation}
Here  $t_n$ is the time of nucleation of a bubble in the Hubble volume: $\Gamma(t_n)/H(t_n)^4=1$ with $H(t)$ being the Hubble parameter. 
By using $\Gamma(t)$, the average spatial fraction of the false vacuum at $t$ is~\cite{Turner:1992tz}
\begin{equation}
\label{eq:fraction}
F(t) = \exp\left[ -\frac{4\pi}{3}\int^t_{t_i} dt' \Gamma(t') a^3(t) r^3(t,t') \right],
\end{equation}
where $t_i$ is the nucleation time of a first bubble, $a(t)$ is the scale factor normalized to unity at the initial time ($t=t_i$), and $r(t,t')$ is the comoving radius of the true vacuum from nucleated time $t'$ to an earlier time $t$, which is given by
\begin{equation}
r(t,t') \equiv \int^t_{t'}  \frac{1}{ a(\tilde{t})}d\tilde{t}.
\end{equation}
Here, we assume that the bubble wall velocity is closed to the light speed.
To evaluate $F(t)$, we need to follow the scale factor evolution by the Friedmann equation:
\begin{equation}
\label{H}
H^2 =  \left(\frac{1}{ a}\frac{da}{ dt} \right)^2=\frac{1}{3} (\rho_v+\rho_r+\rho_w),
\end{equation}
 where $\rho_v$, $\rho_r$ and $\rho_w$ are the energy densities of the vacuum, radiation and the bubble wall, respectively.
 Here, we take the unit of $M_{\rm pl}=1$.
 The bubble wall can be treated as radiation, since the bubble wall velocity is assumed to be the speed of light.
 Thus, the total radiation energy density is given by 
\begin{equation}
\rho_R = \rho_r+\rho_w.
\end{equation}
Evolution of $\rho_R$ can be described by 
\begin{equation}
\label{evo}
\frac{d\rho_R}{dt}+4H\rho_R = -\frac{d\rho_v}{dt}.
\end{equation}
 On the other hand, $ \rho_v$ can be evaluated as 
\begin{equation}
\label{eq:rho_v}
 \rho_v(t)\equiv F(t) \Delta V,
 \end{equation}
 where $\Delta V$ is the difference in the potential energy density between false and true vacua, and we assume that the potential energy density is zero at the true vacuum. 
 Also, $\Delta V$ can be represented by the normarized released latent heat $\alpha$ at the nucleation time, $\alpha = \Delta V/\rho_r$. 
By using $\alpha$, the energy densities of radiation and the (false) vacuum at the initial time (the nucleation time of a first bubble) $t_i$  are given by 
\begin{equation}
\label{eq:rho_inittial}
\rho_R (t_i) = \frac{1}{1+\alpha} \, \rho_{\rm tot} (t_i),
\qquad
\rho_v  (t_i) = \frac{\alpha}{1+\alpha} \, \rho_{\rm tot} (t_i),
\end{equation}
where $\rho_{\rm tot} (t_i) = \rho_R (t_i) + \rho_v (t_i) $ is the total energy density at $t_i$. These values are used as the initial conditions for our numerical calculations, which can affect the PBH abundances as we argue below.


From Eqs.~(\ref{eq:fraction}), (\ref{H}) and (\ref{evo}), we can evaluate the time evolutions of $\rho_v(t)$, $\rho_R(t)$ and $a(t)$.
We then calculate the probability that some Hubble volume collapses into a PBH as 
\begin{equation}
\label{prob}
P(t_n) = \exp\left[ -\frac{4\pi}{3}\int^{t_n}_{t_i} \frac{a^3(t)}{a^3(t_{\rm PBH})} \frac{1}{H^3(t_{\rm PBH})} \Gamma(t) \right],
\end{equation}
where $t_{\rm PBH}$ is the time of production of the PBH, which can be obtained when the density contrast between inside and outside of the Hubble volume exceeds the critical value $\delta_c=0.45$~\cite{Musco:2004ak,Harada:2013epa}.  Such a large energy contrast arises from the probabilistic nature of vacuum decays. 
To evaluate the energy contrast, for fixed values of $\alpha$ and $\beta / H$, we have followed the evolutions of $\rho_R$ and $\rho_v$ for the outside and inside the bubbles by numerically evaluating Eqs.~\eqref{evo} and \eqref{eq:rho_v} with the delayed initial time $t_i$ for the region inside.  By denoting the total energy density outside and inside the bubbles by $\rho_{\rm out}$ and $\rho_{\rm in}$, respectively, the energy density contrast between inside and outside the bubble is given by $\delta = | (\rho_{\rm in} - \rho_{\rm out})|/\rho_{\rm out}$.  $t_{\rm PBH}$ is determied by the time when $\delta$ surpasses $\delta_c$.  We choose the delayed initial time $t_i$ for the region inside to give the maximum $\delta$ for each  $\alpha$ and $\beta / H$.
Once the energy fluctuation exceeds $\delta_c$, the horizon mass can gravitationally collapse to a PBH. 
 The typical mass of PBHs formed at $t_{\rm PBH}$ is  
\begin{equation}
\label{PBHmass}
M_{\rm PBH} \sim \frac{4\pi}{3}H^{-3} (t_{\rm PBH}) \rho_c = 4\pi H^{-1} (t_{\rm PBH}).
\end{equation}
For the case of EWPT, we obtain $M_{\rm PBH}^{\rm EW} \sim 10^{-5} M_{\odot}$, where $M_{\odot}$ is the solar mass.
The mass fraction of such PBHs, $f_{\rm PBH}$, 
in dark matter density can be probed by exploring microlensing effects in observations such as Subaru HSC, OGLE, PRIME and Roman telescope.
The fraction of PBHs from first-order phase transition is generally given by
\begin{equation}
\label{PBHfraction}
f_{\rm PBH} = \left(\frac{H(t_{\rm PBH})}{H(t_0)} \right)^2 \left(\frac{a(t_{\rm PBH})}{a(t_0)} \right)^3 P(t_n) \frac{1}{\Omega_{\rm CDM}},
\end{equation}
where $\Omega_{\rm CDM}$ is the energy density of cold dark matter normalized by the total energy density and $t_0$ is the present time.
For the first-order EWPT, the fraction of PBHs is given by
\begin{equation}
\label{PBHfractionEWPT}
f_{\rm PBH}^{\rm EW} \sim1.49 \times 10^{11} \left( \frac{0.25}{\Omega_{\rm CDM}} \right) \left( \frac{T_{\rm PBH}}{100{\rm GeV}} \right) P(t_n),
\end{equation}
where $T_{\rm PBH}$ is the temperature at the production of PBHs. 

\begin{figure}[t]
  \begin{center}
\includegraphics[width=0.6\textwidth]{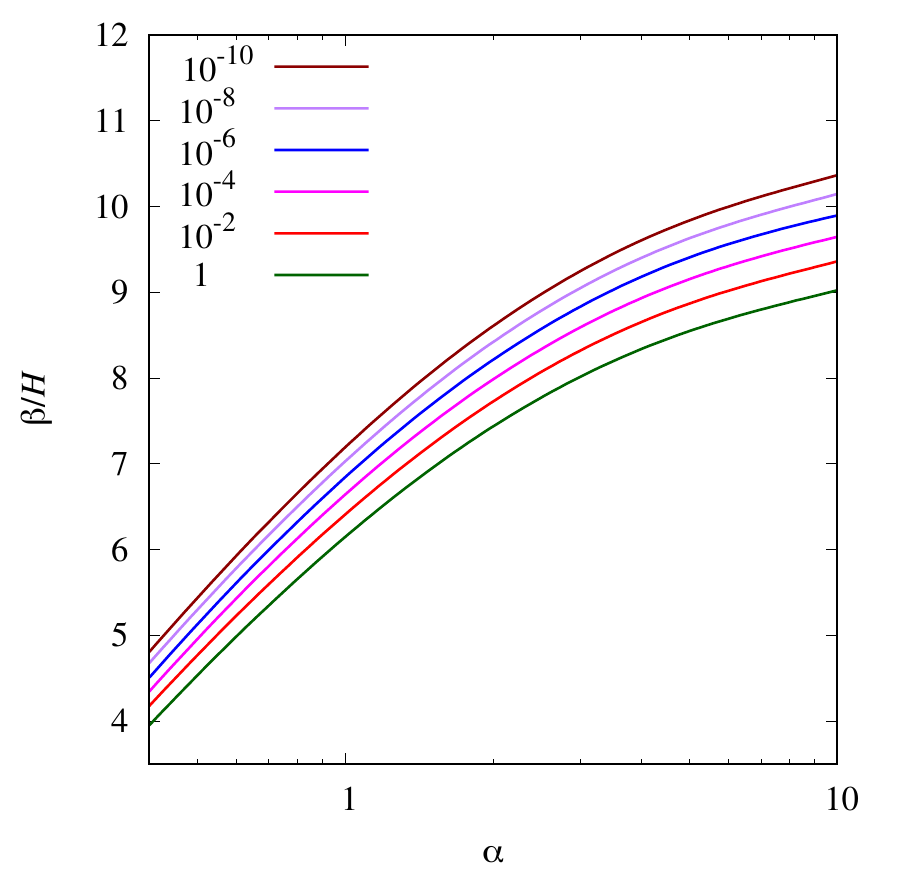}
\caption{ The fraction of PBHs $f_{\rm PBH}$ with respect to $\alpha$ and $\beta/H$ parameters. 
The colored solid lines represent that $f_{\rm PBH}$ = $10^{-10}$ (brown), $10^{-8}$ (purple), $10^{-6}$ (blue), $10^{-4}$ (magenta), $10^{-2}$ (red) and $1$ (green).}
\label{abH}
  \end{center}
\end{figure}

From these equations, $f_{\rm PBH} $ is determined by $P(t_n)$ in Eq.~(\ref{prob}), which can be fixed by $\alpha$ and $\beta$.
In Fig.~\ref{abH}, $f_{\rm PBH}$ is shown as a function of $\alpha$ and $\beta/H$. 
The colored curves are contours $f_{\rm PBH}$ = $10^{-10}$ (brown), $10^{-8}$ (purple), $10^{-6}$ (blue), $10^{-4}$ (magenta), $10^{-2}$ (red) and $1$ (green).
From this figure, $f_{\rm PBH}$ is sensitive to the $\beta/H$ parameter, because $f_{\rm PBH}$ exponentially depends on $\beta/H$.
 In the parameter region below the green curve, PBHs are overproduced.

The GW spectrum from first-order phase transition can also be described by $\alpha$ and $\beta$.
Although three sources, bubble collision (bubble wall kinetic energy), compressional waves (sound waves) and turbulence, can contribute to the GW signal, compressional waves would give a dominant one, and hence we only consider the GWs generated from compressional waves below.
When the bubble wall velocity is closed to the light speed, the fitting function for the GW from compressional waves of plasma is given by~\cite{Hindmarsh:2017gnf,Ellis:2018mja,Caprini:2019egz}
	\begin{align}
	\label{CompGW}
  \Omega_{\rm comp} (f) 
  =   2.061 F_{gw,0}  \left(\frac{\kappa_v\alpha}{1+\alpha}\right)^2  (H(T_n)R_*)
   \tilde{\Omega}_{GW} \left(\frac{f}{\tilde{f}_{\rm comp}}\right)^3
  \left(\frac{7}{4+3(f/\tilde{f}_{\rm comp})^2}\right)^{7/2},  
	  \end{align}
where $F_{gw,0} = 3.57 \times 10^{-5} \left(100/g^t_\ast\right)^{1/3}$ 
with $g_*^t$ being the relativistic degree of freedom, 
$\tilde{\Omega}_{GW}=1.2 \times 10^{-2}$, the mean bubble separation $R_*=v_b(8\pi)^{1/3}/\beta$ 
with $v_b$ being wall velocity and $f_{\rm comp}$ is the peak frequency given by
	\begin{align}
  \tilde{f}_{\rm comp} \simeq 26 \left(\frac{1}{H(T_n)R_*}\right) 
  \left(\frac{T_n}{100~{\rm GeV}}\right)
  \left(\frac{g^t_\ast}{100}\right)^{1/6} 10^{-6}~{\rm Hz}. 
	\end{align}
Here $T_n$ is nucleation temperature, which can be obtained by $\Gamma/H^4|_{T=T_n}=1$.
The expression for the efficiency factor $\kappa_v$ in Eq.~(\ref{CompGW}) can be found in Ref.~\cite{Espinosa:2010hh}.
The GW spectrum  given in Eq.~\eqref{CompGW} should be modified when the sound wave period is shorter than the Hubble rate~\cite{Caprini:2019egz}.
When $H(T_n)R_*>\sqrt{\frac{3}{4}\kappa_v\alpha/(1+\alpha)}$, the GW spectrum is modified as 
	\begin{align}
	\label{CompGW2}
  \Omega_{\rm comp} (f) 
  =   2.061 F_{gw,0}  \left(\frac{\kappa_v\alpha}{1+\alpha}\right)^{3/2}  (H(T_n)R_*)^2
   \tilde{\Omega}_{GW} \left(\frac{f}{\tilde{f}_{\rm comp}}\right)^3
  \left(\frac{7}{4+3(f/\tilde{f}_{\rm comp})^2}\right)^{7/2}. 
	  \end{align}
In order to discuss the testability of the model at GW observation, we used the signal-to-noise ratio
\begin{align}\label{eq:SNR}
\mathrm{SNR}\equiv \sqrt{\delta\times t_{obs} \int^{f_{max}}_{f_{min}}df \left[\frac{h^2 \Omega_{GW}(f)}{h^2\Omega_{\rm sen}(f)}\right]^2},
\end{align}
where $\delta$ is number of independent channel for experiment, $\Omega_{\rm sen}(f)$ is the sensitivity of the GW detector, $t_{obs}$ corresponds to the observation period. 
When the ratio is larger than 10, we could typically detect the GW spectrum~\cite{Caprini:2015zlo}.

 \section{The testability of a specific model}

We here consider what kind of models of electroweak symmetry breaking can produce PBHs at the first-order EWPT.
To this end, we employ the Higgs model with dimension 6 and 8 operators, and examine the parameter space where sufficient amount of PBHs are generated.
We then discuss testability of the model by using PBHs at current and future microlensing observations, in addition to the use of the GWs and the $hhh$ coupling.

It is known that strongly first-order EWPT can be realized in such a model~\cite{Grojean:2004xa}. 
In our analysis, higher order operators in the effective field theory are taken up to dimension 8 operators, because it may be difficult to realize the first-order EWPT only with dimension 6 operators in Ref.~\cite{Postma:2020toi}.
The Higgs potential is given by
\begin{equation}
\label{eq:VEFT}
V_{EFT}(\Phi) =  \mu^2 |\Phi|^2+ \lambda |\Phi|^4 + a_6 |\Phi|^6 + a_8 |\Phi|^8 ,
\end{equation}
where $\Phi$ is the SM-like Higgs doublet field, and $a_8$ is assumed to be positive so that the vacuum is stable.
The coefficients $a_6$ and $a_8$ are parameterized by a dimensionful parameter $\Lambda$ and a dimensionless parameter $\epsilon$, as $a_6=\epsilon/\Lambda^2$, $a_8=1/\Lambda^4$.
We define the classical field of $\Phi$ as  $\langle \Phi^T \rangle=(0,\varphi/\sqrt{2})$.

We consider the effective potential at the one-loop level in the $\overline{\rm MS}$ scheme, which is given by~\cite{Dolan:1973qd}
\begin{align}
\label{eq:Veff}
V_{\rm eff}\left(\varphi, T\right) &= -\frac{\mu^2}{2}\varphi^2+ \frac{\lambda}{4}\varphi^4 + \frac{\epsilon}{8\Lambda^2}\varphi^6 + \frac{1}{16\Lambda^4}\varphi^8\nonumber\\
&\quad+\sum_i\frac{n_i}{64\pi^2} \, M^4_i\left(\varphi\right)\,\left(  \ln\left( \frac{M^2_i\left(\varphi\right)}{Q^2} \right) - c_i \right) + \Delta V_T,  
\end{align}
where $i$ represents species of the SM fields, $c_i$ = 3/2 (for the scalar bosons and the fermions) and 5/6 (for the weak bosons). 
By imposing the stationary condition, the vacuum expectation value is determined, and the renormalized mass of the Higgs boson is defined by the second derivative of $V_{\rm eff}(\varphi, 0)$.
 $\Delta V_T$ is one-loop thermal contribution which is written as 
\begin{align} 
\label{eq:FINI}
\Delta V_T&= \frac{T^4}{2\pi^2}
\left\{ \sum_{i={\rm bosons}} n_i  \int_0^\infty d x x^2\ln 
\left[ 1- \exp \left( -\sqrt{x^2+(M_i(\varphi)/T)^2}\right) \right]\right.\nonumber\\
&\left. + \sum_{i = {\rm fermions}} n_i  \int_0^\infty d x x^2\ln 
\left[ 1+ \exp \left( -\sqrt{x^2+(M_i(\varphi)/T)^2}\right) \right] \right\}.
\end{align}
In order to obtain the ring-improved effective potential, we replace the field dependent masses by
\begin{align}
M^2_i\left(\varphi\right)\to M^2_i\left(\varphi, T\right) = M^2_i\left(\varphi\right) + \Pi_i(T).
\end{align}
The ring-improved field dependent masses of the Higgs boson and Nambu-Goldstone bosons are different from those for the SM fields, which are given by 
\begin{align}
M^2_h\left(\varphi,T\right) = &- \mu^2 +  3\lambda \varphi^2 +   \frac{15\epsilon}{4\Lambda^2} \varphi^4 +  \frac{7}{2\Lambda^4} \varphi^6 +T^2\left(  \frac{ \lambda}{2} + \frac{ 3g^2}{16}  + \frac{g'^2}{16} +\frac{y_t^2}{4} \right),\\
M^2_{NGB}\left(\varphi,T\right) = &- \mu^2 +  \lambda \varphi^2 +   \frac{3\epsilon}{4\Lambda^2} \varphi^4 +  \frac{1}{2\Lambda^4} \varphi^6 +T^2\left(  \frac{ \lambda}{2} + \frac{ 3g^2}{16}  + \frac{g'^2}{16} +\frac{y_t^2}{4} \right),
\end{align}
where $g$ ($g'$) is the SU(2) (U(1)) gauge coupling constant, and $y_t$ is the top Yukawa coupling constant.


 We take into account perturbative unitarity and true vacuum condition in this model for later analyses. 
 Bounds from perturbative unitarity come from two body elastic scatterings among longitudinally polarized weak bosons and the Higgs boson~\cite{Lee:1977eg}. 
 In our model, the most stringent bound is given by the eigenstates from the diagonalization of the S-wave amplitude matrix as
	\begin{equation}
	  	\label{perturbative}
	  	 \frac{1}{16\pi}\left( 4\lambda+15a_6v^2+30a_8v^4
	  	 +\sqrt{4\lambda^2+12\lambda a_6v^2+117a_6^2v^4+675a_8^2v^8 + 540a_6a_8v^6 }\right)<\frac{1}{2}.
	\end{equation}
 In addition, we impose the true vacuum condition which guarantees the minimum at $\varphi=246$~GeV to be the global minimum. 
In the following phase transition analysis, these theoretical requirements are taken into account.

\begin{figure}[t]
  \begin{center}
\includegraphics[width=0.6\textwidth]{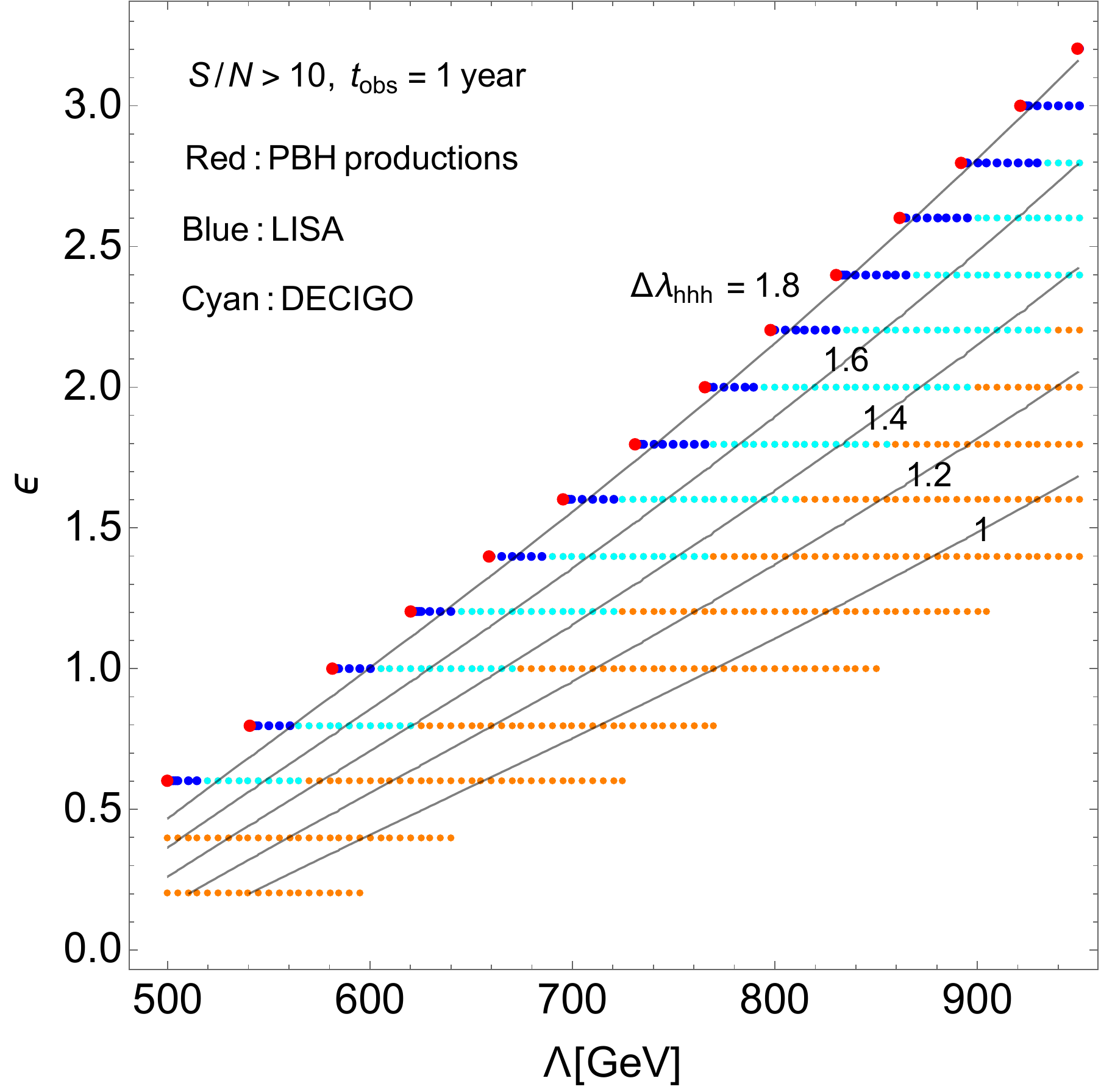}
\caption{ The correlations among the measurements of the $hhh$ coupling, observations of GW spectrum and PBHs with respect to $\epsilon$ and $\Lambda$ parameters.
The grey contours represent the deviations in the $hhh$ coupling from the SM prediction value $\Delta \lambda_{hhh}=$1--1.8 from below.
At the cyan (blue) points, the SNR for the GW from compressional wave of plasma at DECIGO (LISA) experiment with $t_{obs}$ = 1 year is larger than 10.
At the red points with $\alpha\sim0.5$ and $\beta/H<10$, the GW spectrum can be observed at DECIGO and LISA, and the fraction of PBHs can reach the sensitivity of future observations $f_{\rm PBH}>10^{-4}$. 
 }
\label{hGP}
  \end{center}
\end{figure}

We first evaluate the phase transition parameters $\alpha$ and $\beta$ to explore the parameter region where PBHs can be produced.
We have investigated parameters which realize the PBH production criterion of $\delta_c>0.45$.
When the criterion is met, we evaluate the fraction of PBH using Eq.~(\ref{PBHfractionEWPT}).
We have found parameter region where a sufficient fraction of PBH can be produced, which is detectable at current and future microlensing experiments.
It turns out that in order to exceed the criterion relatively large values of $\alpha$ ($\sim 0.5$--$10$) and very small values of $\beta/H$ ($< 10$) are required. 
Furthermore, fraction of PBH is found to be very sensitive to $\beta$.


 In Fig.~\ref{hGP}, the red points where $\alpha\sim0.5$ and $\beta/H<10$ correspond to $f_{\rm PBH}>10^{-4}$, where produced PBHs are expected to be detected by future microlensing observations such as Roman telescope~\cite{Roman2}. 
 Regions where sufficient number of PBHs can be generated are so sensitive to $\beta$.
If we take a slightly larger value for $\Lambda$ than the red points, we get  $\beta/H> O(10)$.
Thus $f_{\rm PBH}$ is too small to be observed or PBHs cannot be produced from first-order EWPT.
In the left region from the red points, the supercooled phase transition may be realized and the linear approximation for the decay rate in Eq.~(\ref{eq:decay}) may not be justified. 
The analysis of PBH production of such case will be a subject of the future work.
The blue and cyan points represents the parameters in which the SNR for the GW spectrum at LISA~\cite{Klein:2015hvg} and DECIGO~\cite{Yagi:2011wg} with $t_{obs}$ = 1 year is larger than 10, respectively.
Regions of the orange points can realize the first-order EWPT.
The grey contours correspond to the deviation in the $hhh$ coupling from the SM prediction, which is defined by 
\begin{equation}
\label{eq:Dhhh}
\Delta\lambda_{hhh}\equiv \frac{\lambda_{hhh}-\lambda_{hhh}^{\rm SM}}{\lambda_{hhh}^{\rm SM}}, \quad \lambda_{hhh}\equiv \left.\frac{\partial^3V_{\rm eff}(\varphi,0)}{\partial \varphi^3}\right|_{\varphi=v}.
\end{equation}

Regions of $f_{\rm PBH}>10^{-2}$ are inside the red points in Fig.~\ref{hGP} which are already in the reach of current observations at Subaru HSC~\cite{Niikura:2017zjd} and OGLE~\cite{Niikura:2019kqi}, and further at PRIME and Roman telescope in the near future.
In these regions, we can test the nature of EWPT by using the PBH observation, in addition to GWs at LISA and DECIGO and the $hhh$ coupling at HL-LHC and ILC.

In this letter, we have employed the effective field theory with dimension 6 and 8 operators.
 Physics of the strongly first-order EWPT basically relies on the non-decoupling property of new physics beyond the SM.
Our analysis in this letter will be extended using the more appropriate non-linear Higgs effective field theory~\cite{Kanemura:2021fvp} elsewhere~\cite{future}.

 \section{Summary}

In this letter, we have evaluated the abundance of PBHs which are produced at the first-order EWPT in the effective field theory with the dimension 6 and 8 operators.
The PBHs produced at the EWPT have masses of about $10^{-5}$ solar mass. 
We have studied parameters of the model which satisfy the condition of PBH generation to exceed the criterion $\delta_c=0.45$.
We have found that there are parameter regions where a sufficient number of PBHs is produced with the fraction $f_{\rm PBH}>10^{-4}$, which can be tested by current and future microlensing observations, such as Subaru HSC, OGLE, PRIME and Roman telescope.
Our results would suggest that PBHs can be used as a new probe of models of the first-order EWPT with a considerablly large duration length of the phase transition and a relatively large latent heat.
In such a case, they can be complementary with measurements of the $hhh$ coupling at future collider experiments and observations of GWs at future space based interferometers. 
Finally, we emphasize that  observations of PBHs with $10^{-5}$ solar mass are performed much earlier than LISA, HL-LHC and also the ILC.
Therefore, we may be able to obtain first information on the nature of EWPT much earlier via PBHs.


\begin{acknowledgments}

We would like to thank Prof.~T.~Sumi for useful discussions. 
The work of S.~K. was supported by the Grant-in-Aid on Innovative Areas, the Ministry of Education, Culture, Sports, Science and Technology, No.~16H06492, and by the JSPS KAKENHI Grant No.~20H00160.
The work of T.~T. was supported by JSPS KAKENHI Grant Number 17H01131, 19K03874 and MEXT KAKENHI Grant Number 19H05110.
\end{acknowledgments}


\end{document}